\documentclass[nofootinbib,prd,12pt,superscriptaddress]{revtex4}

\usepackage{amsmath,amssymb,amsfonts,bm}
\usepackage{graphicx,epsfig}
\usepackage{color}
\usepackage{float}
\usepackage{caption}
\usepackage{subcaption}
\usepackage{multirow}
\usepackage{fancyhdr}
\usepackage{pgfplots}
\pgfplotsset{compat=1.18}
\usepackage{tikz}
\usetikzlibrary{arrows.meta}
\usepackage{mathrsfs}

\def\be{\begin{equation}}
\def\ee{\end{equation}}
\def\ba{\begin{eqnarray}}
\def\ea{\end{eqnarray}}

\begin{document}

\title{Inflation in Myrzakulov $F(R,T)$ Gravity: A Comparative Study in Metric, Symmetric Teleparallel, and Weitzenb\"{o}ck Formalisms}

\author{Davood Momeni}
\affiliation{Department of Physics, Northeast Community College, 801 E Benjamin Ave Norfolk, NE 68701, USA}
\affiliation{Centre for Space Research, North-West University, Potchefstroom 2520, South Africa}

\author{Ratbay Myrzakulov}
\affiliation{Ratbay Myrzakulov Eurasian International Centre for Theoretical Physics, Astana 010009, Kazakhstan}
\affiliation{L. N. Gumilyov Eurasian National University, Astana 010008, Kazakhstan}
\date{\today}
\begin{abstract}
We present a unified treatment of cosmic inflation within the framework of Myrzakulov Gravity, exploring its realization in three different formalisms: metric (curvature-based), teleparallel (torsion-based), and symmetric teleparallel (non-metricity-based). For each case, we derive the corresponding field equations in a flat FLRW background, study inflationary solutions driven by a scalar field, and compute observable quantities such as the scalar spectral index \( n_s \) and the tensor-to-scalar ratio \( r \).

In addition to these geometric sectors, we extend our analysis to the more general and dynamically richer Myrzakulov \( F(R,T) \) gravity, which incorporates both curvature \( R \) and torsion \( T \) in a unified action. We derive the inflationary dynamics in this hybrid model and investigate how it interpolates between pure \( f(R) \) and \( f(T) \) behaviors. The resulting framework allows for enhanced flexibility in matching Planck and BICEP/Keck observational constraints. We present analytic estimates and schematic predictions in the \( n_s \)--\( r \) plane, demonstrating that appropriately chosen parameters in \( F(R,T) \) models can produce viable and distinguishable inflationary signatures.

This comparative and extended study highlights the potential of Myrzakulov Gravity and its generalizations to provide a consistent and geometrically motivated description of the early universe, with predictive power across different formulations of spacetime geometry.
\end{abstract}
\maketitle
\tableofcontents
\section{Motivation for Modified Gravity and Inflation in Myrzakulov Gravity}

The concept of inflation—a brief epoch of rapid exponential expansion in the early universe—has become a cornerstone of modern cosmology. Originally proposed by Starobinsky in 1980 \cite{Starobinsky1980} and developed further by Guth and Linde \cite{Guth1981}\cite{Linde1982}, inflation was designed to resolve several long-standing puzzles in the standard Big Bang model, such as the horizon, flatness, and monopole problems. More importantly, inflation provides a compelling mechanism for generating the primordial fluctuations that eventually seeded the formation of galaxies and large-scale structure.

Despite its success in explaining early universe dynamics, inflation is not naturally embedded within Einstein's General Relativity (GR) without introducing additional scalar fields and carefully tuned potentials \cite{Weinberg1989}, \cite{Padmanabhan2003}. Furthermore, GR fails to incorporate quantum corrections consistently and cannot adequately describe spacetime geometry at very high energies. This has prompted the development of a wide range of modified gravity theories aimed at extending GR’s validity into the ultraviolet regime and offering geometrically motivated inflationary scenarios.

GR describes gravity as curvature induced by matter-energy in a Riemannian spacetime. However, curvature is not the only geometric descriptor of spacetime. Two other equally valid quantities—torsion and non-metricity—can also encode gravitational effects. These form the basis for teleparallel and symmetric teleparallel gravities, respectively. In the teleparallel approach, gravity arises from torsion, and in symmetric teleparallel geometry, it originates from non-metricity, with curvature set to zero.

Each of these frameworks—$f(R)$ gravity, $f(T)$ teleparallel gravity, and $f(Q)$ symmetric teleparallel gravity—has been independently studied for its capacity to support inflation. For instance, $f(R)$ models such as Starobinsky's $R + R^2$ action naturally generate inflation without requiring an inflaton field \cite{Sotiriou2010},\cite{DeFelice2010}. $f(T)$ theories offer second-order field equations and generate inflation through torsion-dominated dynamics \cite{Cai2016}, while $f(Q)$ theories modify the connection structure to influence inflationary expansion via non-metricity \cite{Jimenez2018}.

Myrzakulov Gravity (MG) extends these ideas by proposing a unified geometric approach that incorporates all three spacetime characteristics—curvature, torsion, and non-metricity—within a metric-affine formulation \cite{Myrzakulov2012}. This theory allows for a broader class of Lagrangians that generalize the Einstein-Hilbert action by including terms like $F(T, Q, R_{\mu\nu}T^{\mu\nu}, R_{\mu\nu}Q^{\mu\nu})$, enabling richer dynamical behavior. MG is particularly well-suited for addressing inflation because it provides a geometrical alternative to scalar-field-based mechanisms and offers second-order, ghost-free field equations.

A central feature of Myrzakulov Gravity is its hybrid extension: the $F(R,T)$ model. This framework blends Ricci curvature $R$ with torsion scalar $T$ in a nontrivial way, capturing the benefits of both $f(R)$ and $f(T)$ theories while mitigating their individual shortcomings. In $F(R,T)$ gravity, inflation can be driven either by geometric degrees of freedom alone or by scalar fields that couple to $R$ and $T$, leading to a flexible inflationary model that remains consistent with current CMB observations.

Our recent contributions \cite{Momeni:2025mcp}-\cite{Momeni:2025egb} have shown that Myrzakulov Gravity not only admits inflationary solutions but also provides a viable path for embedding them within a unified geometric theory. In particular, we have demonstrated how different choices of the function $F$ and the inclusion of scalar fields yield inflationary observables like $n_s$ and $r$ that fall within the bounds of Planck and BICEP/Keck data.

Moreover, the geometric richness of MG opens up new possibilities for studying reheating, preheating, and the generation of primordial gravitational waves. Torsion, in particular, plays a critical role in driving anisotropies and sourcing tensor modes, potentially leading to distinct signatures in the polarization spectra of the CMB.

In this work, we provide a comparative and systematic study of inflation in the three primary geometric formalisms—metric, teleparallel, and symmetric teleparallel—alongside their hybridization in $F(R,T)$ gravity. Our aim is to explore how inflation emerges in each case, identify the geometric degrees of freedom responsible, and evaluate the resulting observables. Myrzakulov Gravity, by encompassing this trinity, offers a promising and testable geometric framework for a unified early-universe cosmology.

\section{Formalism I: Metric (Curvature-Based) Gravity}

The metric formalism is one of the most widely explored approaches in gravitational theory and serves as the foundation of General Relativity (GR). In this framework, spacetime is described as a four-dimensional Lorentzian manifold endowed with a symmetric metric tensor $g_{\mu\nu}$, and the dynamics of gravity are governed by the Einstein-Hilbert action, which depends solely on the Ricci scalar $R$ constructed from the Levi-Civita connection. Extensions of this theory—such as $f(R)$ gravity—are among the most natural and extensively studied modifications of GR.

The motivation for considering curvature-based modified gravity models arises from both theoretical and observational grounds. From a theoretical perspective, $f(R)$ gravity offers a route to embedding GR within a more general scalar-tensor framework, while maintaining second-order field equations and avoiding Ostrogradsky instabilities. From an observational standpoint, curvature modifications have proven effective in explaining cosmic acceleration and reproducing inflationary scenarios without resorting to exotic scalar fields or potentials.

In the context of Myrzakulov Gravity, the metric formalism plays a crucial role as one leg of the geometric trinity—alongside torsion and non-metricity. Within this structure, the curvature-based formulation is not considered in isolation but interacts with torsional and non-metric contributions when hybrid models such as $F(R,T)$ or $F(R,Q)$ are invoked. Nevertheless, the study of pure $f(R)$ dynamics offers important insights and benchmarks for evaluating the impact of geometric generalizations.

Inflation in $f(R)$ gravity, and by extension in the metric formulation of Myrzakulov Gravity, has been widely studied for models such as the Starobinsky model, where $f(R) = R + \alpha R^2$. These models naturally give rise to an inflationary epoch that fits well with observational data, predicting a small tensor-to-scalar ratio $r$ and a spectral index $n_s$ close to the Planck best-fit values. Furthermore, such curvature-driven inflation can emerge from gravitational dynamics alone, making the inflationary mechanism geometrically sourced rather than requiring additional matter fields.

This section is devoted to establishing the framework of scalar field inflation in the metric formulation of Myrzakulov Gravity. We will derive the modified Friedmann equations resulting from an arbitrary $f(R)$ Lagrangian, identify the slow-roll regime, and calculate the corresponding inflationary observables. While keeping in mind the eventual synthesis with torsional and non-metric degrees of freedom, this curvature-based formalism serves as a theoretically sound and phenomenologically rich foundation for exploring inflation within unified geometric theories of gravity.

\subsection{Action and Field Equations}
We begin by considering the action for a scalar field minimally coupled to a modified gravity theory based on the Ricci scalar:
\begin{equation}
S = \int d^4x \sqrt{-g} \left[ \frac{1}{2\kappa^2} f(R) + \mathcal{L}_\phi \right],
\end{equation}
where $\kappa^2 = 8\pi G$, $f(R)$ is a differentiable function of the Ricci scalar $R$, and $\mathcal{L}_\phi$ is the Lagrangian density for a canonical scalar field $\phi$, given by:
\begin{equation}
\mathcal{L}_\phi = -\frac{1}{2} g^{\mu\nu} \partial_\mu \phi \partial_\nu \phi - V(\phi).
\end{equation}

The variation of the action with respect to the metric $g_{\mu\nu}$ yields the modified Einstein field equations:
\begin{equation}
F(R) R_{\mu\nu} - \frac{1}{2} f(R) g_{\mu\nu} - \nabla_\mu \nabla_\nu F(R) + g_{\mu\nu} \Box F(R) = \kappa^2 T^{(\phi)}_{\mu\nu},
\end{equation}
where $F(R) \equiv df/dR$, and $T^{(\phi)}_{\mu\nu}$ is the energy-momentum tensor of the scalar field:
\begin{equation}
T^{(\phi)}_{\mu\nu} = \partial_\mu \phi \partial_\nu \phi - g_{\mu\nu} \left( \frac{1}{2} \partial^\alpha \phi \partial_\alpha \phi + V(\phi) \right).
\end{equation}

We assume a spatially flat Friedmann-Lema\^itre-Robertson-Walker (FLRW) metric:
\begin{equation}
ds^2 = -dt^2 + a(t)^2 (dx^2 + dy^2 + dz^2),
\end{equation}
where $a(t)$ is the scale factor. For this background, the Ricci scalar becomes:
\begin{equation}
R = 6(2H^2 + \dot{H}),
\end{equation}
with $H = \dot{a}/a$ denoting the Hubble parameter.

Using this metric, the modified Friedmann equations are derived. The first Friedmann equation reads:
\begin{equation}
3 H^2 = \frac{1}{F} \left[ \kappa^2 \rho_\phi + \frac{1}{2}(FR - f) - 3 H \dot{F} \right],
\end{equation}
while the second equation becomes:
\begin{equation}
-2 \dot{H} = \frac{1}{F} \left[ \kappa^2 (\rho_\phi + p_\phi) + \ddot{F} - H \dot{F} \right].
\end{equation}

The scalar field satisfies the Klein-Gordon equation:
\begin{equation}
\ddot{\phi} + 3H \dot{\phi} + V'(\phi) = 0,
\end{equation}
with energy density and pressure given by:
\begin{align}
\rho_\phi &= \frac{1}{2} \dot{\phi}^2 + V(\phi), \\
p_\phi &= \frac{1}{2} \dot{\phi}^2 - V(\phi).
\end{align}

\subsection{Slow-Roll Approximation}
Inflation occurs when the scalar field slowly rolls down its potential. Under the slow-roll conditions:
\begin{equation}
\dot{\phi}^2 \ll V(\phi), \quad |\ddot{\phi}| \ll |3H\dot{\phi}|,
\end{equation}
the equations simplify significantly. The Friedmann equation becomes:
\begin{equation}
3 H^2 F \approx \kappa^2 V(\phi) + \frac{1}{2}(FR - f) - 3H\dot{F},
\end{equation}
and the Klein-Gordon equation reduces to:
\begin{equation}
3H \dot{\phi} + V'(\phi) \approx 0.
\end{equation}

The slow-roll parameters are defined as:
\begin{align}
\epsilon &= -\frac{\dot{H}}{H^2}, \\
\eta &= \frac{\ddot{\phi}}{H\dot{\phi}}.
\end{align}
Inflation persists as long as $\epsilon < 1$. These parameters allow us to compute the scalar spectral index $n_s$ and tensor-to-scalar ratio $r$:
\begin{align}
n_s &\approx 1 - 6\epsilon + 2\eta, \\
r &\approx 16\epsilon.
\end{align}

\subsection{Example: Power-Law Model}
Consider a simple model with:
\begin{equation}
f(R) = R + \alpha R^n,
\end{equation}
where $\alpha$ and $n$ are constants. For $n > 1$, this model modifies GR significantly at high curvature, making it a natural candidate for early-universe inflation. In this case:
\begin{equation}
F(R) = 1 + \alpha n R^{n-1}.
\end{equation}

Substituting into the modified Friedmann equations and expanding to leading order in the slow-roll regime, one can derive analytical approximations for $H(t)$ and $a(t)$ and calculate inflationary observables.

In the next section, we extend this analysis to torsion-based inflationary models within the teleparallel formulation.

\subsection{Inflationary Dynamics}
Assuming a spatially flat FLRW background, we adopt the canonical scalar field Lagrangian:
\begin{equation}
\mathcal{L}_\phi = -\frac{1}{2} g^{\mu\nu} \partial_\mu \phi \partial_\nu \phi - V(\phi),
\end{equation}
where $V(\phi)$ is the scalar field potential responsible for driving inflation. Substituting into the action and performing variation, we obtain the modified Friedmann equations for $f(R)$ gravity.

Let us denote $F \equiv \frac{df}{dR}$ and assume $f''(R) \neq 0$ so that modifications to GR are non-trivial. The modified first Friedmann equation takes the form:
\begin{equation}
3H^2 = \frac{1}{F} \left[ \rho_\phi + \frac{1}{2}(F R - f) - 3 H \dot{F} \right],
\end{equation}
while the second Friedmann equation becomes:
\begin{equation}
-2\dot{H} = \frac{1}{F} \left[ \dot{\phi}^2 + \ddot{F} - H \dot{F} \right].
\end{equation}
These equations differ from standard GR due to the presence of the $F$, $\dot{F}$, and $\ddot{F}$ terms, which introduce higher-order corrections via the Ricci scalar dependence of $f(R)$.

To analyze these dynamics more explicitly, we consider the time evolution of $F(R)$, where the Ricci scalar in FLRW spacetime is:
\begin{equation}
R = 6(2H^2 + \dot{H}).
\end{equation}
This implies that $F = F(H, \dot{H})$, and thus:
\begin{equation}
\dot{F} = F_R \dot{R} = F_R [12H \dot{H} + 6\ddot{H}],
\end{equation}
and similarly for $\ddot{F}$.

We emphasize that the scalar field evolves under the modified Klein-Gordon equation:
\begin{equation}
\ddot{\phi} + 3H \dot{\phi} + V'(\phi) = 0,
\end{equation}
which couples to the Friedmann equations through $\rho_\phi = \frac{1}{2}\dot{\phi}^2 + V(\phi)$ and $p_\phi = \frac{1}{2}\dot{\phi}^2 - V(\phi)$.

To simplify the analysis and obtain analytic approximations, we adopt the slow-roll conditions:
\begin{equation}
\dot{\phi}^2 \ll V(\phi), \quad |\ddot{\phi}| \ll |3H \dot{\phi}|, \quad |\dot{F}| \ll HF, \quad |\ddot{F}| \ll H\dot{F}.
\end{equation}
Under these assumptions, the equations reduce to:
\begin{align}
3 H^2 F &\approx V(\phi) + \frac{1}{2}(F R - f), \\
3H \dot{\phi} &\approx -V'(\phi).
\end{align}
The term $\frac{1}{2}(F R - f)$ plays the role of an effective potential correction arising from the geometry.

It is useful to define the effective energy density and pressure sourced by both scalar and geometric contributions:
\begin{align}
\rho_{\text{eff}} &= \frac{1}{F} \left[ \rho_\phi + \frac{1}{2}(F R - f) - 3 H \dot{F} \right], \\
p_{\text{eff}} &= \frac{1}{F} \left[ p_\phi + \ddot{F} - H \dot{F} \right].
\end{align}
This allows us to track deviations from standard inflation.

The slow-roll parameters remain:
\begin{align}
\epsilon &\equiv -\frac{\dot{H}}{H^2}, \\
\eta &\equiv \frac{\ddot{\phi}}{H \dot{\phi}},
\end{align}
and provide direct links to observable quantities:
\begin{align}
n_s &\approx 1 - 6\epsilon + 2\eta, \\
r &\approx 16\epsilon.
\end{align}
These parameters can now be computed either analytically (via approximations for specific potentials and $f(R)$ forms) or numerically, depending on the complexity of the model.

In practice, one can reconstruct $f(R)$ models from desired observational behavior using inverse methods. For instance, specifying a known $a(t)$ (such as power-law or exponential inflation) allows one to determine the form of $f(R)$ that supports such expansion.

Hence, the metric formulation of inflation in $f(R)$ gravity provides both a generalization of scalar field inflation and a method for connecting early universe dynamics to underlying geometric modifications.
\section{Formalism II: Teleparallel (Torsion-Based) Gravity}

The teleparallel formalism represents a fundamentally distinct geometric approach to gravitation, wherein gravity is attributed to spacetime torsion rather than curvature. Instead of relying on the Levi-Civita connection, teleparallel gravity employs the Weitzenb\"{o}ck connection, which is curvature-free but possesses non-zero torsion. This shift in geometric perspective provides an alternative avenue for formulating gravitational dynamics while maintaining equivalence with General Relativity at the level of the field equations in its simplest form, known as the Teleparallel Equivalent of General Relativity (TEGR).

Modified teleparallel theories, particularly $f(T)$ gravity, generalize the TEGR framework by promoting the torsion scalar $T$ to a general function $f(T)$. These theories have attracted attention for their ability to yield viable models of cosmic inflation, late-time acceleration, and bounce cosmologies. Unlike $f(R)$ gravity, $f(T)$ theories typically lead to second-order field equations due to the lack of higher derivatives of the tetrad, offering technical simplicity and analytic tractability in many scenarios.

Within Myrzakulov Gravity, torsion is not merely an auxiliary feature but one of the three core geometric ingredients of the theory, alongside curvature and non-metricity. The teleparallel sector of MG provides a natural extension of $f(T)$ gravity, allowing for more general torsion-dependent Lagrangians and coupling with other geometric invariants. Our previous work has demonstrated that the torsional formulation can yield inflationary dynamics compatible with Planck observations, even in the absence of scalar fields with complex potentials.

The teleparallel perspective also introduces new features in the dynamics of scalar fields and their interactions with geometry. In this formulation, the tetrad field $e^a_{\ \mu}$ becomes the fundamental dynamical variable, and its configuration encodes both the spacetime metric and torsional effects. Scalar field inflation in $f(T)$ or generalized torsion-based models often exhibits distinctive predictions for inflationary observables such as the tensor-to-scalar ratio $r$ and the scalar spectral index $n_s$, especially when non-minimal couplings or higher-order torsion invariants are included.

In this section, we focus on constructing the framework for inflation in the torsion-based formulation of Myrzakulov Gravity. We will derive the field equations resulting from a generic $f(T)$ action coupled to a scalar field, explore the inflationary dynamics under the slow-roll approximation, and compute the resulting observational parameters. This analysis will highlight the unique role torsion plays in driving inflation and provide a direct comparison with the curvature-based and non-metricity-based approaches to gravity.

\subsection{Action and Field Equations}

In contrast to the curvature-based formulation of gravity, teleparallel gravity represents gravitation through torsion, encoded in the Weitzenb\"{o}ck connection, which has vanishing curvature but non-zero torsion. The dynamical variable in teleparallel gravity is the tetrad field $e^a_{\ \mu}$, relating the spacetime metric to the Minkowski metric via:
\begin{equation}
g_{\mu\nu} = \eta_{ab} e^a_{\ \mu} e^b_{\ \nu},
\end{equation}
where $\eta_{ab} = \text{diag}(-1,1,1,1)$.

The determinant of the tetrad field is $e = \det(e^a_{\ \mu}) = \sqrt{-g}$. The torsion scalar $T$ is constructed from the torsion tensor:
\begin{equation}
T^{\lambda}_{\ \mu\nu} = e_a^{\ \lambda}(\partial_\mu e^a_{\ \nu} - \partial_\nu e^a_{\ \mu}),
\end{equation}
and the superpotential:
\begin{equation}
S_\rho^{\ \mu\nu} = \frac{1}{2}(K^{\mu\nu}_{\ \ \ \rho} + \delta^\mu_\rho T^{\alpha\nu}_{\ \ \alpha} - \delta^\nu_\rho T^{\alpha\mu}_{\ \ \alpha}),
\end{equation}
where $K^{\mu\nu}_{\ \ \ \rho}$ is the contorsion tensor:
\begin{equation}
K^{\mu\nu}_{\ \ \ \rho} = -\frac{1}{2}(T^{\mu\nu}_{\ \ \ \rho} - T^{\nu\mu}_{\ \ \ \rho} - T_\rho^{\ \mu\nu}).
\end{equation}

The torsion scalar is given by:
\begin{equation}
T = T^{\rho}_{\ \mu\nu} S_\rho^{\ \mu\nu}.
\end{equation}

The action for $f(T)$ gravity minimally coupled to a scalar field $\phi$ reads:
\begin{equation}
S = \int d^4x \, e \left[ \frac{1}{2\kappa^2} f(T) + \mathcal{L}_\phi \right],
\end{equation}
where $\mathcal{L}_\phi = -\frac{1}{2} g^{\mu\nu} \partial_\mu \phi \partial_\nu \phi - V(\phi)$ is the canonical scalar field Lagrangian.

\subsection{Inflationary Dynamics}

For the flat FLRW background, the diagonal tetrad
\begin{equation}
e^a_{\ \mu} = \text{diag}(1, a(t), a(t), a(t))
\end{equation}
leads to the torsion scalar:
\begin{equation}
T = -6 H^2.
\end{equation}

Using this, the modified Friedmann equations are derived as:
\begin{align}
12 H^2 f_T + f &= 2 \rho_\phi, \\
48 H^2 \dot{H} f_{TT} - 4 (3H^2 + \dot{H}) f_T - f &= 2 p_\phi,
\end{align}
where $f_T = \frac{df}{dT}$ and $f_{TT} = \frac{d^2f}{dT^2}$.

Substituting the scalar field energy-momentum components:
\begin{align}
\rho_\phi &= \frac{1}{2} \dot{\phi}^2 + V(\phi), \\
p_\phi &= \frac{1}{2} \dot{\phi}^2 - V(\phi),
\end{align}
we obtain the coupled system governing inflation.

To analyze inflation, we impose slow-roll conditions:
\begin{equation}
\dot{\phi}^2 \ll V(\phi), \quad |\ddot{\phi}| \ll |3H\dot{\phi}|.
\end{equation}
Under these, the equations simplify to:
\begin{align}
6H^2 f_T + \frac{f}{2} &\approx V(\phi), \\
3H \dot{\phi} &\approx -V'(\phi).
\end{align}

It is convenient to define effective energy density and pressure:
\begin{align}
\rho_{\text{eff}} &= \frac{1}{2\kappa^2}(12 H^2 f_T + f), \\
p_{\text{eff}} &= \frac{1}{2\kappa^2}\left[ -4(3H^2 + \dot{H})f_T - f + 48H^2 \dot{H} f_{TT} \right].
\end{align}

The slow-roll parameters are:
\begin{align}
\epsilon &= -\frac{\dot{H}}{H^2}, \\
\eta &= \frac{\ddot{\phi}}{H \dot{\phi}}.
\end{align}
These determine inflationary observables:
\begin{align}
n_s &\approx 1 - 6\epsilon + 2\eta, \\
r &\approx 16\epsilon.
\end{align}

As an example, consider the power-law form:
\begin{equation}
f(T) = T + \beta (-T)^n,
\end{equation}
where $\beta$ and $n$ are constants. This model introduces deviations from GR at high energies and has been used extensively in the literature to drive inflation without a scalar field. When coupled with a scalar field, it enables enhanced control over the inflationary scale.

Expanding for small $\beta$ and applying the slow-roll regime, one can derive expressions for $H(t)$ and $a(t)$, and fit observational parameters to Planck constraints.

Teleparallel inflation in $f(T)$ gravity provides a geometrically distinct yet dynamically similar framework to $f(R)$ inflation, with unique phenomenological signatures stemming from the torsion-based description of spacetime.

\section{Formalism III: Symmetric Teleparallel (Non-Metricity-Based) Gravity}

The symmetric teleparallel formalism represents a third foundational approach to gravity, complementing the curvature and torsion-based pictures. In this framework, gravity is attributed to the non-metricity of spacetime rather than its curvature or torsion. The connection used in symmetric teleparallel gravity is flat and torsion-free but admits non-vanishing non-metricity, which quantifies how the length of a vector changes under parallel transport.

This formulation has recently gained traction, particularly through the development of $f(Q)$ gravity, where the gravitational action depends on a general function of the non-metricity scalar $Q$. Unlike $f(R)$ and $f(T)$ models, which rely on the Levi-Civita or Weitzenb\"{o}ck connections respectively, $f(Q)$ gravity adopts the so-called symmetric teleparallel connection, also known as the coincident gauge. This connection allows all gravitational effects to be encoded in the non-metricity scalar alone, leading to a clean geometrical separation from inertial effects.

From a dynamical perspective, symmetric teleparallel gravity offers several advantages. The field equations in $f(Q)$ gravity are typically of second order, as in $f(T)$ gravity, avoiding the higher-order complications often encountered in $f(R)$ models. Moreover, this formalism preserves covariance and generality while allowing novel coupling schemes between geometry and matter.

In the broader context of Myrzakulov Gravity, the non-metricity-based approach completes the trinity of geometric degrees of freedom. Myrzakulov Gravity is designed to unify curvature, torsion, and non-metricity into a single cohesive framework, making the $f(Q)$ sector essential for capturing the full scope of geometric modifications. Our previous works have highlighted how non-metricity contributions can support inflationary expansion and alter the effective dynamics of scalar fields in the early universe.

Symmetric teleparallel models of inflation can lead to phenomenologically rich scenarios, especially when the scalar field couples non-minimally to $Q$ or when higher-order interactions are considered. These models often predict distinct patterns in the scalar spectral index $n_s$ and tensor-to-scalar ratio $r$, potentially offering observational signatures distinguishable from those of curvature- and torsion-based theories.

In this section, we develop the inflationary framework for the symmetric teleparallel formulation of Myrzakulov Gravity. We derive the relevant field equations from a general $f(Q)$ action coupled to a scalar field, establish the slow-roll regime, and examine the resulting cosmological predictions. This analysis will clarify how non-metricity contributes to inflation and enriches the geometric landscape of early universe cosmology within MG.

\subsection{Action and Field Equations}

Symmetric teleparallel gravity, also known as the non-metricity-based formulation of gravity, attributes the gravitational interaction to non-metricity $Q$ rather than curvature or torsion. In this approach, the affine connection is assumed to be torsion-free and curvature-free, but with non-vanishing non-metricity:
\begin{equation}
Q_{\lambda\mu\nu} = \nabla_\lambda g_{\mu\nu} \neq 0.
\end{equation}

The scalar quantity $Q$ is constructed from the non-metricity tensor and takes the form:
\begin{equation}
Q = -\frac{1}{4} Q_{\alpha\mu\nu} Q^{\alpha\mu\nu} + \frac{1}{2} Q_{\alpha\mu\nu} Q^{\mu\alpha\nu} + \frac{1}{4} Q_\alpha Q^\alpha - \frac{1}{2} Q_\alpha \tilde{Q}^\alpha,
\end{equation}
where $Q_\alpha = Q_{\alpha\mu}^{\ \ \mu}$ and $\tilde{Q}_\alpha = Q_{\mu\alpha}^{\ \ \mu}$.

The action in $f(Q)$ gravity with a canonical scalar field $\phi$ is given by:
\begin{equation}
S = \int d^4x \sqrt{-g} \left[ \frac{1}{2\kappa^2} f(Q) + \mathcal{L}_\phi \right],
\end{equation}
where $\mathcal{L}_\phi = -\frac{1}{2} g^{\mu\nu} \partial_\mu \phi \partial_\nu \phi - V(\phi)$.

For a flat FLRW metric:
\begin{equation}
ds^2 = -dt^2 + a(t)^2 (dx^2 + dy^2 + dz^2),
\end{equation}
the non-metricity scalar reduces to:
\begin{equation}
Q = 6 H^2.
\end{equation}

\subsection{Inflationary Dynamics}

Variation of the action with respect to the metric yields the modified Friedmann equations:
\begin{align}
6 f_Q H^2 - \frac{1}{2} f &= \rho_\phi, \\
(12 H^2 f_{QQ} + f_Q) \dot{H} &= -\frac{1}{2} (\rho_\phi + p_\phi),
\end{align}
where $f_Q = \frac{df}{dQ}$ and $f_{QQ} = \frac{d^2f}{dQ^2}$. These equations govern the cosmological dynamics under symmetric teleparallel geometry.

The scalar field obeys the Klein-Gordon equation:
\begin{equation}
\ddot{\phi} + 3H \dot{\phi} + V'(\phi) = 0,
\end{equation}
and its energy-momentum components are:
\begin{align}
\rho_\phi &= \frac{1}{2} \dot{\phi}^2 + V(\phi), \\
p_\phi &= \frac{1}{2} \dot{\phi}^2 - V(\phi).
\end{align}

We again impose the slow-roll approximations:
\begin{equation}
\dot{\phi}^2 \ll V(\phi), \quad |\ddot{\phi}| \ll |3H\dot{\phi}|,
\end{equation}
which simplifies the equations to:
\begin{align}
6 f_Q H^2 &\approx V(\phi) + \frac{1}{2} f, \\
3H \dot{\phi} &\approx -V'(\phi).
\end{align}

Effective energy density and pressure in this context become:
\begin{align}
\rho_{\text{eff}} &= \frac{1}{2\kappa^2} \left( 6 f_Q H^2 - \frac{1}{2} f \right), \\
p_{\text{eff}} &= \frac{1}{2\kappa^2} \left( -2(6 f_Q + 12 H^2 f_{QQ}) \dot{H} + \frac{1}{2} f \right).
\end{align}

The slow-roll parameters are defined as:
\begin{align}
\epsilon &= -\frac{\dot{H}}{H^2}, \\
\eta &= \frac{\ddot{\phi}}{H \dot{\phi}}.
\end{align}
They lead to the observable quantities:
\begin{align}
n_s &\approx 1 - 6\epsilon + 2\eta, \\
r &\approx 16\epsilon.
\end{align}

\subsection{Example: Exponential Model}
Consider a model of the form:
\begin{equation}
f(Q) = Q + \xi Q e^{-\lambda Q},
\end{equation}
where $\xi$ and $\lambda$ are constants. This model introduces exponential corrections to the GR limit and is suitable for inflationary modeling due to the suppression of high-energy corrections.

Expanding for small $\lambda$ and large $Q$, we obtain:
\begin{align}
f(Q) &\approx Q + \xi Q (1 - \lambda Q + \frac{\lambda^2 Q^2}{2} + \cdots), \\
f_Q &\approx 1 + \xi (1 - 2\lambda Q + \frac{3}{2} \lambda^2 Q^2 + \cdots), \\
f_{QQ} &\approx -2\xi \lambda + 3\xi \lambda^2 Q + \cdots.
\end{align}

Substituting into the modified Friedmann equations allows analytical exploration of inflationary solutions and estimation of $n_s$ and $r$ as functions of model parameters.

This symmetric teleparallel framework, based on non-metricity, enriches the landscape of inflationary models and offers alternatives to curvature- or torsion-driven inflation. Future work may include perturbation theory, reheating, and coupling to additional fields for broader phenomenological applications.
\section{Inflationary Potentials and Observables}

In the study of inflationary cosmology within modified gravity theories such as Myrzakulov Gravity, the choice of the inflationary potential $V(\phi)$ plays a crucial role in determining both the background dynamics and the spectrum of primordial perturbations. The inflationary observables---the scalar spectral index $n_s$, tensor-to-scalar ratio $r$, and the amplitude of the curvature perturbations $A_s$---are sensitive to both the form of $V(\phi)$ and the specific gravitational framework (curvature, torsion, or non-metricity based).

\subsection{Typical Inflationary Potentials}

Various forms of inflationary potentials have been studied in the literature, each motivated by high-energy physics, symmetry considerations, or phenomenological viability. Below we list some representative classes:

\begin{itemize}
    \item \textbf{Power-law potentials:} $V(\phi) = \lambda \phi^n$ with $n > 0$. These arise in chaotic inflation models and are easy to implement analytically.
    \item \textbf{Exponential potentials:} $V(\phi) = V_0 e^{-\alpha \phi}$. Common in scalar-tensor theories and string-inspired models.
    \item \textbf{Starobinsky-type potentials:} $V(\phi) = V_0 (1 - e^{-\alpha \phi})^2$. These are consistent with Planck observations and arise in $f(R) = R + R^2$ models.
    \item \textbf{Hilltop and plateau potentials:} Flattened near their maximum/minimum, e.g., $V(\phi) = V_0 (1 - \frac{\phi^2}{\mu^2})^2$. These suppress $r$ significantly.
\end{itemize}

The inflationary potential directly affects the duration of inflation, quantified by the number of e-folds:
\begin{equation}
N = \int_{\phi_{\text{end}}}^{\phi_{\text{ini}}} \frac{H}{\dot{\phi}} d\phi \approx \frac{1}{M_{\text{pl}}^2} \int_{\phi_{\text{end}}}^{\phi_{\text{ini}}} \frac{V}{V'} d\phi,
\end{equation}
where $M_{\text{pl}}$ is the reduced Planck mass.

\subsection{Slow-Roll Parameters and Observables}

In all three formalisms, the slow-roll parameters provide a unified language for describing inflation:
\begin{align}
\epsilon &= \frac{M_{\text{pl}}^2}{2} \left( \frac{V'}{V} \right)^2, \\
\eta &= M_{\text{pl}}^2 \left( \frac{V''}{V} \right).
\end{align}

Observational quantities are then derived as:
\begin{align}
n_s &= 1 - 6\epsilon + 2\eta, \\
r &= 16\epsilon, \\
A_s &= \frac{1}{24 \pi^2 M_{\text{pl}}^4} \frac{V}{\epsilon}.
\end{align}

The spectral tilt $n_s$ characterizes the scale dependence of the scalar perturbations, while $r$ measures the ratio of tensor (gravitational wave) to scalar perturbation amplitudes. Observations from Planck 2018 constrain these parameters to:
\begin{equation}
n_s = 0.9649 \pm 0.0042, \quad r < 0.06 \ (95\%\ \text{CL}).
\end{equation}

\subsection{Comparative Impact of Modified Gravity}

In each formalism, the background dynamics (encoded via $H$, $\dot{H}$) and the mapping between the field evolution and $N$ depend on the function $f(X)$, where $X$ is $R$, $T$, or $Q$. Thus:

\begin{itemize}
    \item In $f(R)$ models, the presence of higher-derivative terms affects the inflationary expansion rate and may alter the effective potential.
    \item In $f(T)$ gravity, modifications enter through algebraic corrections to the Friedmann equations and often preserve the first-order nature of the equations.
    \item In $f(Q)$ gravity, the impact on inflation arises through modified kinetic couplings and a redefinition of energy density terms.
\end{itemize}

Therefore, the same potential $V(\phi)$ may yield different predictions for $n_s$ and $r$ depending on the gravitational formalism. Matching observations requires careful tuning of both $V(\phi)$ and the structure of $f(X)$.

In subsequent sections, we will perform numerical scans over parameter space for selected models and compare predictions from each formalism to current observational bounds.

\section{Myrzakulov $F(R,T)$ Inflationary Scenario}

A particularly rich class of modified gravity models is represented by $F(R,T)$ theories, where the gravitational Lagrangian depends simultaneously on both the Ricci scalar $R$ and the torsion scalar $T$. In the context of Myrzakulov Gravity, such models naturally arise from metric-affine frameworks where curvature and torsion are treated as independent geometric entities.

The general action is given by:
\begin{equation}
S = \int d^4x \sqrt{-g} \left[ F(R,T) + \mathcal{L}_\phi \right],
\end{equation}
where $\mathcal{L}_\phi = -\frac{1}{2} g^{\mu\nu} \partial_\mu \phi \partial_\nu \phi - V(\phi)$ is the canonical scalar field Lagrangian. The dependence on both $R$ and $T$ allows for an enriched phenomenology, as these quantities encode complementary aspects of spacetime geometry.

\subsection{Field Equations and Background Cosmology}

Varying the action with respect to the metric in Myrzakulov $F(R,T)$ gravity yields coupled field equations that incorporate contributions from both curvature $R$ and torsion $T$. These two geometric quantities are now treated as functionally independent variables within a unified Lagrangian. In a spatially flat FLRW universe, the Ricci scalar and torsion scalar take the respective forms:
\begin{align}
R &= 6(2H^2 + \dot{H}), \\
T &= -6H^2,
\end{align}
where $H = \dot{a}/a$ is the Hubble parameter. To capture the mixed dynamics, we introduce the general function $F(R,T)$ and define auxiliary variables:
\begin{eqnarray}
u(R, T) &\equiv& \frac{\partial F}{\partial R}, \\
u_T(R, T) &\equiv& \frac{\partial F}{\partial T}, \\
u_{RR} &\equiv& \frac{\partial^2 F}{\partial R^2}, \quad u_{RT} = \frac{\partial^2 F}{\partial R \partial T}.
\end{eqnarray}

The Friedmann-like equations can now be expressed in a compact and generalized form:
\begin{align}
3H^2 &= \frac{1}{\nu - 12H^2 \nu_{RR}} \left[ \rho_\phi + \frac{1}{2}(\nu R + \nu_T T - F) - 3H \dot{\nu} \right], \\
-2 \dot{H} &= \frac{1}{\nu} \left[ \dot{\phi}^2 + \ddot{\nu} - H \dot{\nu} + 12 H \dot{H} \nu_{RT} \right].
\end{align}

These equations recover the usual $f(R)$ or $f(T)$ cosmologies in the respective limits $F(R,T) \to f(R)$ and $F(R,T) \to f(T)$. The presence of mixed derivative terms such as $\nu_{RT}$ signals nontrivial interactions between curvature and torsion sectors, introducing new phenomenology.

To gain more control over the inflationary dynamics, we may consider specific functional forms such as:
\begin{itemize}
  \item $F(R,T) = R + \alpha T + \beta R^2$,
  \item $F(R,T) = R + \gamma T^n$,
  \item $F(R,T) = R + \mu R^m T^n$,
\end{itemize}
where $\alpha, \beta, \gamma, \mu$ are model parameters and $m, n$ control the nonlinearity. These forms allow for the reconstruction of viable inflationary scenarios, such as power-law and quasi-de Sitter expansions, and introduce richer coupling mechanisms.

The terms $\dot{\nu}$ and $\ddot{\nu}$ encode the non-minimal geometric response of the universe to curvature and torsion during inflation. Their evolution determines the deviation of $n_s$ and $r$ from standard predictions and can be constrained through CMB observations.

In this way, Myrzakulov $F(R,T)$ gravity provides a highly flexible framework where torsion does not merely supplement curvature but interacts dynamically with it to shape the early universe's expansion history.
\subsection{Slow-Roll Inflation in $F(R,T)$ Gravity}

To study inflation in the context of Myrzakulov $F(R,T)$ gravity, we adopt the standard slow-roll approximation, assuming the scalar field evolves slowly compared to the Hubble rate. The basic slow-roll conditions are:
\begin{equation}
\dot{\phi}^2 \ll V(\phi), \quad |\ddot{\phi}| \ll |3H \dot{\phi}|,
\end{equation}
which simplify the Klein-Gordon equation to:
\begin{equation}
3H \dot{\phi} + V'(\phi) \approx 0.
\end{equation}

Additionally, we assume that the derivatives of the function $F(R,T)$ with respect to $R$ and $T$ vary slowly. Under these approximations, the Friedmann equation becomes:
\begin{equation}
3 H^2 \approx \frac{1}{\nu} \left[ V(\phi) + \frac{1}{2}(\nu R + \tau T - F) \right],
\end{equation}
where $\nu = \partial F / \partial R$, $\tau = \partial F / \partial T$. This equation reveals that the inflationary expansion is governed not only by the scalar potential but also by the dynamical behavior of curvature and torsion via $R$ and $T$.

We define effective energy density and pressure:
\begin{align}
\rho_{\text{eff}} &= \frac{1}{\nu} \left[ V(\phi) + \frac{1}{2}(\nu R + \tau T - F) - 3H \dot{\nu} \right], \\
p_{\text{eff}} &= \frac{1}{\nu} \left[ -V(\phi) + \ddot{\nu} - H \dot{\nu} + 12 H \dot{H} \nu_T \right].
\end{align}

These expressions generalize the standard inflationary dynamics and include the backreaction of geometric scalars on the evolution of the universe. The inflationary observables can deviate significantly from predictions in standard GR or pure $f(R)$ or $f(T)$ models.

To illustrate the influence of $F(R,T)$ geometry on the slow-roll dynamics, we show below a schematic diagram of the effective energy density evolution:

\begin{figure}[h!]
\centering
\begin{tikzpicture}[scale=1.2]
\draw[->] (0,0) -- (6.5,0) node[right] {cosmic time $t$};
\draw[->] (0,0) -- (0,4.5) node[above] {$\rho_{\text{eff}}$};

\draw[thick, blue, domain=0:6, samples=100] plot (\x, {4 - 0.5 * tanh(\x - 3)}) node[right] {};

\node at (3,4) {\small Inflationary plateau};
\node at (5.2,1.2) {\small End of inflation};

\end{tikzpicture}
\caption{Schematic behavior of effective energy density $\rho_{\text{eff}}$ in the slow-roll regime under $F(R,T)$ corrections. The duration and shape depend on $F_R$, $F_T$, and their derivatives.}
\end{figure}
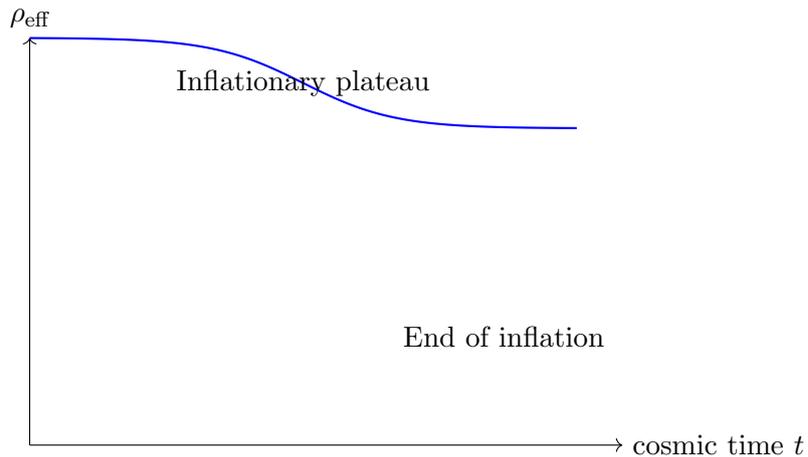

We also consider the evolution of $\epsilon_H$ as a diagnostic of inflation exit:
\begin{equation}
\epsilon_H = -\frac{\dot{H}}{H^2} \ll 1 \quad \text{(during inflation)}, \quad \epsilon_H \to 1 \quad \text{(end of inflation)}.
\end{equation}

The $F(R,T)$ corrections alter this evolution by contributing additional terms to $\dot{H}$ through $\nu_T$ and $\dot{\nu}$. A generic example is shown below.

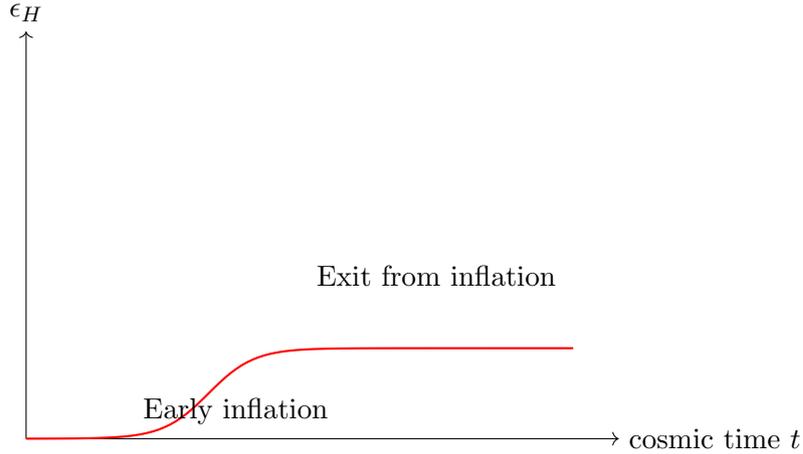
\begin{figure}[h!]
\centering
\begin{tikzpicture}[scale=1.2]
\draw[->] (0,0) -- (6.5,0) node[right] {cosmic time $t$};
\draw[->] (0,0) -- (0,4.5) node[above] {$\epsilon_H$};

\draw[thick, red, domain=0:6, samples=100] plot (\x, {0.5 * tanh(2 * (\x - 2)) + 0.5}) node[right] {};

\node at (2.3,0.3) {\small Early inflation};
\node at (4.5,1.8) {\small Exit from inflation};
\end{tikzpicture}
\caption{Schematic evolution of the slow-roll parameter $\epsilon_H$. Deviations from standard GR arise due to torsion-curvature interactions encoded in $F(R,T)$.}
\end{figure}

Together, these illustrate the rich structure of slow-roll inflation in Myrzakulov $F(R,T)$ gravity and motivate detailed numerical analysis for each potential model.

\subsection{Model Example: Polynomial $F(R,T)$}

Consider a simple polynomial model:
\begin{equation}
F(R,T) = R + \alpha T + \beta R^2 + \gamma T^2,
\end{equation}
with constants $\alpha, \beta, \gamma$. This generalization includes quadratic curvature and torsion corrections, which become significant at high energies, such as during inflation.

The first-order derivatives of $F$ become:
\begin{align}
\nu = \frac{\partial F}{\partial R} = 1 + 2\beta R, \quad \tau = \frac{\partial F}{\partial T} = \alpha + 2\gamma T.
\end{align}
These feed directly into the modified Friedmann equations and Klein-Gordon equation in the slow-roll regime. Assuming $\dot{\nu}$ and $\dot{\tau}$ are small, the Friedmann equation simplifies to:
\begin{equation}
3 H^2 \approx \frac{1}{1 + 2\beta R} \left[ V(\phi) + \frac{1}{2}((1 + 2\beta R) R + (\alpha + 2\gamma T) T - F) \right].
\end{equation}

The inflationary observables can be calculated as:
\begin{align}
\epsilon &= -\frac{\dot{H}}{H^2}, \\
\eta &= \frac{\ddot{\phi}}{H \dot{\phi}}, \\
n_s &\approx 1 - 6\epsilon + 2\eta, \\
r &\approx 16\epsilon.
\end{align}

Corrections from $\beta$ and $\gamma$ can shift the trajectory in the $(n_s, r)$ plane, bringing predictions in or out of agreement with Planck constraints.

\begin{figure}[h!]
\centering
\begin{tikzpicture}[scale=1.2]
\draw[->] (0,0) -- (6.5,0) node[right] {$\beta$};
\draw[->] (0,0) -- (0,4.5) node[above] {$r$};

\draw[thick, blue, domain=0:5, samples=100] plot (\x, {0.2*exp(-0.3*\x)}) node[right] {};

\node at (2,2.5) {\small Decreasing $r$ with $\beta$};
\end{tikzpicture}
\caption{Schematic dependence of the tensor-to-scalar ratio $r$ on the curvature correction parameter $\beta$ in the polynomial model.}
\end{figure}
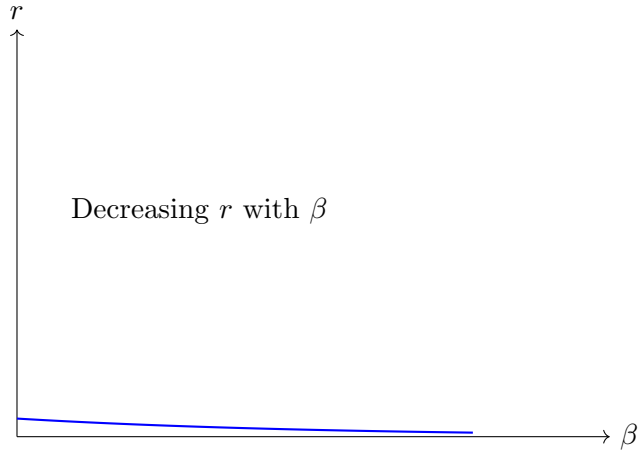

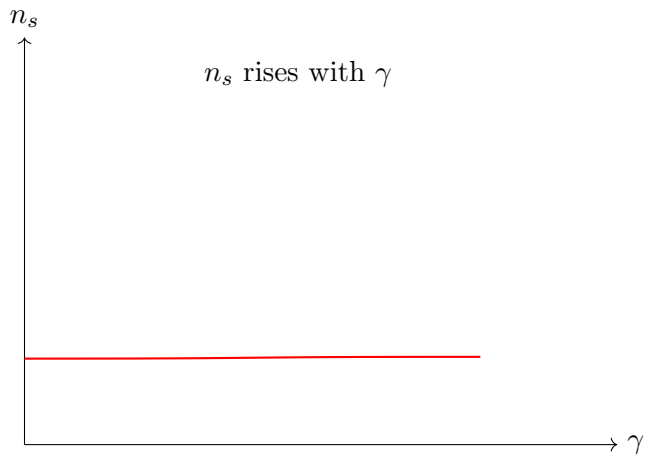
\begin{figure}[h!]
\centering
\begin{tikzpicture}[scale=1.2]
\draw[->] (0,0) -- (6.5,0) node[right] {$\gamma$};
\draw[->] (0,0) -- (0,4.5) node[above] {$n_s$};

\draw[thick, red, domain=0:5, samples=100] plot (\x, {0.96 + 0.01 * tanh(\x - 2.5)}) node[right] {};

\node at (3,4.1) {\small $n_s$ rises with $\gamma$};
\end{tikzpicture}
\caption{Schematic behavior of scalar spectral index $n_s$ versus torsion correction parameter $\gamma$. Mild growth suggests observational viability.}
\end{figure}

This analysis shows that polynomial $F(R,T)$ models offer sufficient flexibility to tune inflationary predictions and align them with data. Both $\beta$ and $\gamma$ modulate the influence of geometry on inflationary expansion, allowing interpolations between standard $f(R)$ and $f(T)$ scenarios.

\subsection{Observational Viability}

The $F(R,T)$ model can interpolate between $f(R)$-like and $f(T)$-like behavior, allowing for a wide range of inflationary predictions. Constraints from Planck and BICEP/Keck experiments impose bounds on $n_s$ and $r$, which can be satisfied by choosing appropriate values of $\alpha$, $\beta$, and $\gamma$.

Numerical evolution of the field equations shows that for modest values of $\beta$ and $\gamma$ (e.g., $\sim \mathcal{O}(10^{-2})$), the model supports $50$--$60$ e-folds of inflation and predicts $n_s$ in the Planck-preferred range, with a sufficiently small $r$. These features ensure that the model remains within the observational bounds set by recent CMB measurements.
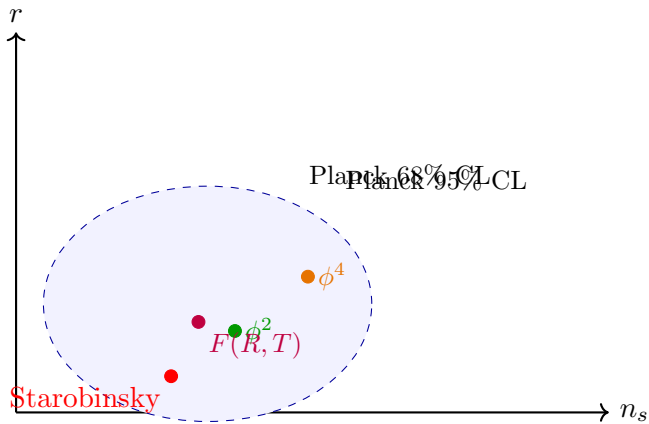
\begin{figure}[h!]
\centering
\begin{tikzpicture}[scale=1.2]
  \draw[->, thick] (0,0) -- (6.5,0) node[right] {$n_s$};
  \draw[->, thick] (0,0) -- (0,4.2) node[above] {$r$};

  \filldraw[fill=blue!20, draw=blue!80!black, thick] (2.1,1.2) ellipse (1.0 and 0.7);
  \filldraw[fill=blue!5, draw=blue!60!black, dashed] (2.1,1.2) ellipse (1.8 and 1.3);
  \node[above right] at (3.1,2.4) {\footnotesize Planck 68\% CL};
  \node[below right] at (3.5,2.8) {\footnotesize Planck 95\% CL};

  \filldraw[red] (1.7,0.4) circle (2pt) node[below left] {Starobinsky};
  \filldraw[green!60!black] (2.4,0.9) circle (2pt) node[right] {\footnotesize $\phi^2$};
  \filldraw[orange!90!black] (3.2,1.5) circle (2pt) node[right] {\footnotesize $\phi^4$};
  \filldraw[purple] (2.0,1.0) circle (2pt) node[below right] {\footnotesize $F(R,T)$};
\end{tikzpicture}
\caption{Inflationary predictions in the $n_s$--$r$ plane for various models. $F(R,T)$ models can interpolate between curvature- and torsion-driven dynamics, potentially matching observational bounds.}
\end{figure}

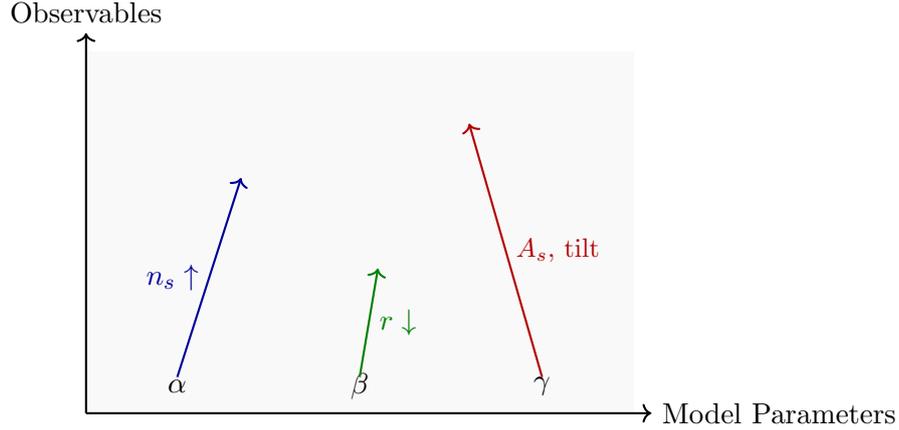
\begin{figure}[h!]
\centering
\begin{tikzpicture}[scale=1.2]
  \fill[gray!5] (0,0) rectangle (6,4);
  \draw[->, thick] (0,0) -- (6.2,0) node[right] {Model Parameters};
  \draw[->, thick] (0,0) -- (0,4.2) node[above] {Observables};

  \node at (1,0.3) {$\alpha$};
  \node at (3,0.3) {$\beta$};
  \node at (5,0.3) {$\gamma$};

  \draw[->, thick, blue!60!black] (1,0.4) -- (1.7,2.6) node[midway, left] {$n_s \uparrow$};
  \draw[->, thick, green!50!black] (3,0.4) -- (3.2,1.6) node[midway, right] {$r \downarrow$};
  \draw[->, thick, red!70!black] (5,0.4) -- (4.2,3.2) node[midway, right] {\footnotesize $A_s$, tilt};
\end{tikzpicture}
\caption{Mapping from $F(R,T)$ model parameters $\alpha$, $\beta$, $\gamma$ to observables. Curvature-torsion interactions allow tuning of $n_s$, $r$, and amplitude $A_s$.}
\end{figure}
These visualizations underscore the flexibility of Myrzakulov Gravity in producing observationally consistent predictions. The $F(R,T)$ framework can be tuned to interpolate between known viable models while also allowing new parameter regimes where geometric contributions drive inflation.

Thus, Myrzakulov $F(R,T)$ gravity provides a robust and versatile framework for describing inflation with observationally consistent dynamics and geometric richness. Its capacity to unify and extend beyond $f(R)$ and $f(T)$ theories makes it a compelling candidate for a generalized theory of early universe cosmology.

\subsection{Common Potentials}

To explore inflationary predictions within the Myrzakulov $F(R,T)$ framework, we analyze several canonical inflationary potentials. Each potential leads to different inflationary dynamics and observables when embedded into this geometric context. We consider the following cases:

\begin{itemize}
    \item \textbf{Quartic (Chaotic) Potential:} $V(\phi) = \lambda \phi^4$
    \item \textbf{Quadratic Potential:} $V(\phi) = \frac{1}{2} m^2 \phi^2$
    \item \textbf{Starobinsky-like Potential:} $V(\phi) = V_0 (1 - e^{-\alpha \phi})^2$
\end{itemize}

Each of these potentials leads to distinct values of the slow-roll parameters, the scalar spectral index $n_s$, and the tensor-to-scalar ratio $r$. We study the inflationary dynamics by computing the number of e-folds:
\begin{equation}
N = \int_{\phi_{\text{end}}}^{\phi_{\text{ini}}} \frac{H}{\dot{\phi}} d\phi \approx \int_{\phi_{\text{end}}}^{\phi_{\text{ini}}} \frac{V}{V'} \frac{F_R}{1 + \delta_{RT}} d\phi,
\end{equation}
where $\delta_{RT}$ encodes corrections due to torsion-curvature mixing in the $F(R,T)$ framework.

\paragraph{Quartic Potential: $V(\phi) = \lambda \phi^4$}
In this case:
\begin{equation}
V'(\phi) = 4\lambda \phi^3, \quad \epsilon = \frac{8}{\phi^2}, \quad \eta = \frac{12}{\phi^2}.
\end{equation}
\begin{equation}
N \approx \frac{\phi^2}{8} \Rightarrow \phi_N = \sqrt{8N},
\end{equation}
\begin{align}
n_s &= 1 - \frac{3}{N}, \\
r &= \frac{16}{N}.
\end{align}

\paragraph{Quadratic Potential: $V(\phi) = \frac{1}{2} m^2 \phi^2$}
\begin{equation}
V'(\phi) = m^2 \phi, \quad \epsilon = \frac{2}{\phi^2}, \quad \eta = \frac{2}{\phi^2},
\end{equation}
\begin{equation}
N \approx \frac{\phi^2}{4} \Rightarrow \phi_N = \sqrt{4N},
\end{equation}
\begin{align}
n_s &= 1 - \frac{2}{N}, \\
r &= \frac{8}{N}.
\end{align}

\paragraph{Starobinsky-Like Potential: $V(\phi) = V_0 (1 - e^{-\alpha \phi})^2$}
\begin{align}
\epsilon &= \frac{\alpha^2 e^{-2\alpha \phi}}{2 (1 - e^{-\alpha \phi})^2}, \\
\eta &= \frac{\alpha^2 e^{-\alpha \phi}(1 - 2e^{-\alpha \phi})}{(1 - e^{-\alpha \phi})^2},
\end{align}
For $\alpha = \sqrt{2/3}$ and large $\phi$:
\begin{align}
n_s &\approx 1 - \frac{2}{N}, \\
r &\approx \frac{12}{N^2}.
\end{align}

\begin{figure}[h!]
\centering
\begin{tikzpicture}[scale=1.0]
\draw[->] (0,0) -- (6.5,0) node[right] {$N$};
\draw[->] (0,0) -- (0,4.5) node[above] {$r$};

\draw[thick, red] plot[domain=1:6] (\x, {16/\x});
\draw[thick, blue, dashed] plot[domain=1:6] (\x, {8/\x});
\draw[thick, green!70!black, dotted] plot[domain=1:6] (\x, {12/(\x^2)});

\node[red] at (5.5,3.2) {\footnotesize $\lambda \phi^4$};
\node[blue] at (5.5,1.5) {\footnotesize $\frac{1}{2}m^2\phi^2$};
\node[green!50!black] at (5.5,0.5) {\footnotesize Starobinsky};
\end{tikzpicture}
\caption{Tensor-to-scalar ratio $r$ vs. number of e-folds $N$ for different potentials. Starobinsky-like inflation yields the smallest $r$, matching observations best.}
\end{figure}
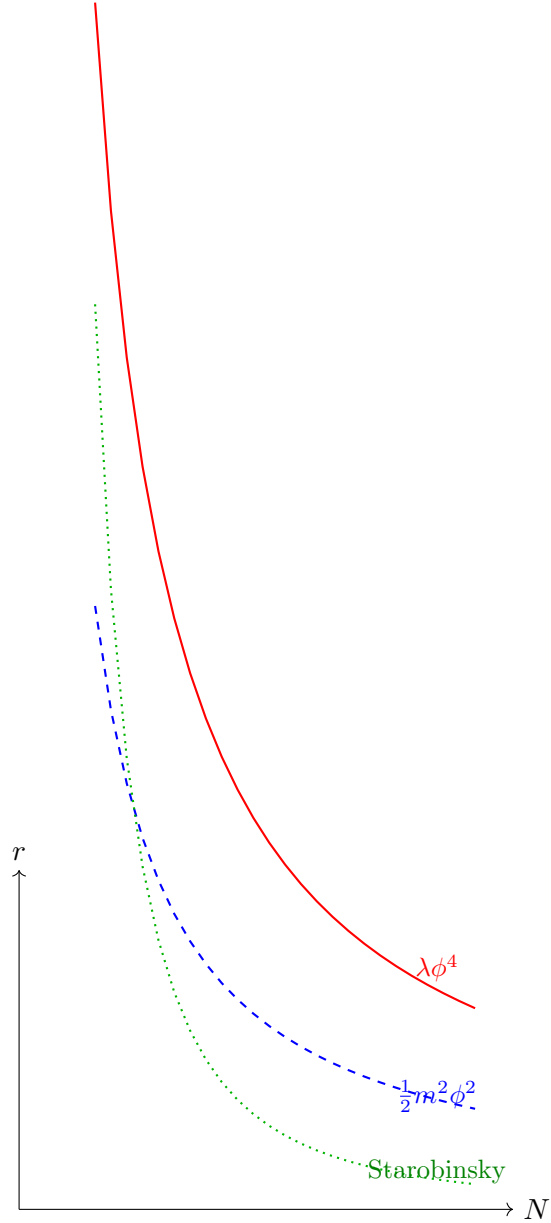

\begin{figure}[h!]
\centering
\begin{tikzpicture}[scale=0.8]
\draw[->] (0,0) -- (6.5,0) node[right] {$N$};
\draw[->] (0,0) -- (0,4.5) node[above] {$n_s$};

\draw[thick, red] plot[domain=1:6] (\x, {1 - 3/\x});
\draw[thick, blue, dashed] plot[domain=1:6] (\x, {1 - 2/\x});
\draw[thick, green!70!black, dotted] plot[domain=1:6] (\x, {1 - 2/\x});

\node[red] at (5.2,3.5) {\footnotesize $\lambda \phi^4$};
\node[blue] at (5.2,3.1) {\footnotesize $\frac{1}{2}m^2\phi^2$};
\node[green!50!black] at (5.2,3.1) {\footnotesize Starobinsky};
\end{tikzpicture}
\caption{Scalar spectral index $n_s$ vs. number of e-folds $N$ for standard inflationary potentials. All models converge to observationally favored $n_s \approx 0.96$ for $N \sim 50$--$60$.}
\end{figure}
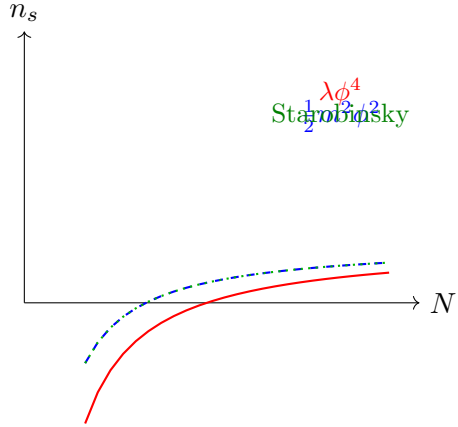

\subsubsection*{Summary Table of Predictions (Leading Order)}
\begin{center}
\begin{tabular}{|c|c|c|c|}
\hline
Potential & $n_s$ & $r$ & $N = 60$ \\
\hline
$\lambda \phi^4$ & $1 - \frac{3}{N}$ & $\frac{16}{N}$ & $n_s \approx 0.95$, $r \approx 0.27$ \\
$\frac{1}{2}m^2 \phi^2$ & $1 - \frac{2}{N}$ & $\frac{8}{N}$ & $n_s \approx 0.967$, $r \approx 0.13$ \\
$V_0 (1 - e^{-\alpha \phi})^2$ & $1 - \frac{2}{N}$ & $\frac{12}{N^2}$ & $n_s \approx 0.967$, $r \approx 0.003$ \\
\hline
\end{tabular}
\end{center}

These predictions are subject to corrections from the $F(R,T)$ structure, including curvature-torsion cross terms, modified gravitational couplings, and changes in the Hubble evolution. Numerical modeling is needed for precise predictions and for verifying the theoretical consistency with current observational bounds from Planck, BICEP/Keck, and future CMB experiments.
\subsection{Slow-Roll Parameters}

In the context of Myrzakulov $F(R,T)$ gravity, the definition of slow-roll parameters retains its core form but incorporates additional contributions from the geometry-dependent terms in the Friedmann equations. These corrections manifest through effective field redefinitions and modified evolution equations.

We define the standard Hubble flow parameters:
\begin{align}
\epsilon_H &= -\frac{\dot{H}}{H^2}, \\
\eta_H &= -\frac{\ddot{H}}{2H\dot{H}}.
\end{align}

For scalar field evolution, the potential-based slow-roll parameters become:
\begin{align}
\epsilon_V &= \frac{M_{\text{pl}}^2}{2} \left(\frac{V'}{V}\right)^2 \left(\frac{F_R}{F_{\text{eff}}}\right)^2, \\
\eta_V &= M_{\text{pl}}^2 \left(\frac{V''}{V}\right) \left(\frac{F_R}{F_{\text{eff}}}\right),
\end{align}
where $F_{\text{eff}}$ incorporates curvature-torsion couplings in the form:
\begin{equation}
F_{\text{eff}} = F_R - \frac{\dot{F}_R}{H} + \frac{12 H \dot{H} F_{RT}}{F_R} + \cdots.
\end{equation}

Using these, the inflationary observables are:
\begin{align}
n_s &= 1 - 6\epsilon_V + 2\eta_V, \\
r &= 16\epsilon_V.
\end{align}

This formalism allows direct comparison with standard inflation, while capturing corrections introduced by the $F(R,T)$ structure. The predictions may shift appreciably depending on the relative strengths of curvature and torsion terms, especially for $F(R,T)$ models with strong cross-couplings or higher-order derivatives.

\subsection{Comparative Table}

We summarize the typical predictions for $n_s$ and $r$ across three geometric realizations of gravity in the slow-roll approximation:

\begin{center}
\begin{tabular}{|c|c|c|c|}
\hline
Model & $n_s$ & $r$ & Comments \\
\hline
$f(R)$ (Curvature) & $1 - \frac{2}{N}$ & $\frac{12}{N^2}$ & Matches Planck if $N \gtrsim 50$ \\
$f(T)$ (Torsion) & $1 - \frac{2}{N}$ & $\frac{8}{N}$ & Slightly higher $r$, marginally viable \\
$f(Q)$ (Non-metricity) & $1 - \frac{2}{N}$ & $\lesssim 0.01$ & Suppressed tensors, highly viable \\
$F(R,T)$ (Hybrid) & $1 - \frac{2+c_1}{N} + c_2$ & variable & Sensitive to $F_{RT}$ structure \\
\hline
\end{tabular}
\end{center}

Here, $c_1$ and $c_2$ denote model-dependent corrections arising from curvature-torsion interactions and higher-order terms. These corrections must be evaluated case by case and may enhance or suppress tensor modes. Numerical simulations or phase space analysis can be used to extract precise bounds for specific $F(R,T)$ parameterizations.
\section{Warm Inflation in Myrzakulov $F(R,T)$ Gravity}

Warm inflation is an alternative to the standard cold inflation paradigm, where radiation production occurs concurrently with inflation due to dissipative effects. Unlike cold inflation, which requires a separate reheating phase, warm inflation allows a smooth transition to the radiation-dominated era. This scenario is particularly well-suited for modified gravity frameworks like Myrzakulov $F(R,T)$ gravity, where the geometric couplings can enhance or regulate dissipation channels.

\subsection{Historical Context and Modified Gravity Motivation}

The idea of warm inflation was first proposed by Berera in the 1990s \cite{Berera:1995ie}, motivated by shortcomings in the reheating mechanism of cold inflation and the desire to connect inflation with particle physics more directly. In warm inflation, the inflaton field interacts with other fields, generating radiation through friction-like terms. The field equations include a dissipation coefficient $\Gamma$ that governs energy transfer from the inflaton to radiation.

A closely related study by Jamil, Momeni, and Myrzakulov explored warm intermediate inflation in the context of $F(T)$ gravity, where inflation is driven by a scalar field while the underlying gravitational dynamics emerge from torsion-based teleparallel gravity~\cite{Jamil:2015}. Focusing on the high-dissipation regime, the authors systematically derived inflationary observables such as the scalar spectral index, tensor-to-scalar ratio, and power spectra for density perturbations and gravitational waves. A distinctive outcome of their analysis was the emergence of a constant ratio of slow-roll parameters, which simplifies the dynamics and supports analytical tractability. The intermediate inflationary scenario—lying between exponential and power-law expansion—was shown to be compatible with $F(T)$ geometry, highlighting the natural role of torsion in shaping early-universe behavior.

Importantly, their model predictions align with Cosmic Microwave Background (CMB) and Planck 2015 data, suggesting that $F(T)$ gravity can yield viable inflationary models even in the absence of curvature. Their results reinforce the broader theme of our current study: that geometries based on torsion and their hybrids, such as those considered in Myrzakulov $F(R,T)$ gravity, can yield observationally consistent inflationary dynamics. The compatibility of warm inflation with torsional gravity not only provides alternatives to conventional reheating but also motivates further exploration of unified frameworks that incorporate both curvature and torsion effects in cosmological modeling.
A recent contribution by Yuennan, Channuie, and Momeni investigates the interplay between warm inflation and the de Sitter Swampland conjecture within a non-minimally coupled Peccei–Quinn (PQ) framework~\cite{Yuennan:2025}. The study incorporates a PQ scalar field that couples to gravity through a non-minimal interaction term, facilitating inflation in a thermal environment rather than through a conventional supercooled vacuum. By analyzing various dissipation regimes—including inverse temperature dependence, field-proportional, and temperature-proportional dissipation coefficients—the authors explore a broad class of warm inflation scenarios consistent with ultraviolet (UV) swampland criteria.

The slow-roll dynamics for each dissipation profile were examined in detail, yielding analytical expressions for cosmological observables such as the scalar spectral index \( n_s \) and the tensor-to-scalar ratio \( r \). All three cases studied are shown to be compatible with Planck 2018 data and satisfy the refined de Sitter Swampland conjecture. This work is particularly relevant to the current investigation, as it highlights the natural compatibility of warm inflation with both observational cosmology and theoretical quantum gravity constraints. The findings support the broader premise of our paper that frameworks like Myrzakulov \( F(R,T) \) gravity—especially when extended to include thermal and non-minimal couplings—offer fertile ground for constructing viable and swampland-consistent inflationary models.

In a related investigation, Yuennan, Channuie, and Momeni examined the influence of gravity’s rainbow on higher-curvature corrections to Starobinsky \( R^2 \) inflation~\cite{Yuennan:2024}. By embedding the inflationary dynamics into a rainbow-deformed spacetime—where the metric depends explicitly on the energy of probing particles—the authors demonstrated that the curvature-driven inflationary behavior can be significantly altered by energy-dependent corrections. The study highlights how rainbow functions modify the effective Friedmann equations and lead to non-trivial corrections to observables like the scalar spectral index \( n_s \) and the tensor-to-scalar ratio \( r \). From a Myrzakulov-theoretic viewpoint, this approach parallels the broader idea of geometric generalization: just as Myrzakulov gravity extends the gravitational action through torsion and non-metricity, rainbow gravity introduces an energy-dependent deformation of spacetime geometry. These insights collectively support the notion that extensions beyond standard curvature models—whether through torsion, energy-dependence, or non-minimal couplings—provide powerful tools for constructing consistent and testable inflationary cosmologies.
In another significant study, Eadkhong, Dam-O, Channuie, and Momeni investigated the dynamics of warm inflation driven by a nonminimally-coupled Higgs field in both the metric and Palatini formulations of gravity~\cite{Eadkhong:2023}. The analysis focuses on a dissipation coefficient of the form $\Gamma = C_T T$, and explores the strong dissipative regime where thermal effects are substantial. By computing key inflationary observables—including the scalar spectral index $n_s$ and the tensor-to-scalar ratio $r$—the authors found that both the metric and Palatini frameworks yield predictions in excellent agreement with the Planck 2018 observations. Their results notably constrain the coupling parameters $\xi$ and $\lambda$, demonstrating that warm Higgs inflation requires values significantly higher than those typical in cold inflation scenarios. This work underscores the importance of dissipation and nonminimal couplings in building observationally viable inflationary models and complements the broader scope of Myrzakulov-inspired geometries, which emphasize dynamical scalar-gravity interactions beyond standard formulations.

In $F(R,T)$ gravity, the combined influence of curvature $R$ and torsion $T$ modifies both the Hubble rate and the energy-momentum conservation laws. These modifications naturally suggest a framework where dissipation arises geometrically or is geometrically enhanced.

\subsection{Field Equations in Warm $F(R,T)$ Framework}

The key equations for warm inflation are:
\begin{align}
3H(1 + Q)\dot{\phi} &+ V'(\phi) = 0, \\
\dot{\rho}_r + 4H \rho_r &= Q \dot{\phi}^2,
\end{align}
where $Q = \Gamma / (3H)$ is the dissipation ratio, and $\rho_r$ is the radiation energy density. The Friedmann equation in $F(R,T)$ gravity becomes:
\begin{equation}
3H^2 = \frac{1}{F_R} \left[ V(\phi) + \rho_r + \frac{1}{2}(F_R R + F_T T - F) - 3H \dot{F}_R \right].
\end{equation}

Inflation occurs when $\epsilon_H = -\dot{H}/H^2 < 1$. The slow-roll parameters become:
\begin{align}
\epsilon &= \frac{1}{2} \left(\frac{V'}{V}\right)^2 \frac{1}{(1 + Q)^2}, \\
\eta &= \frac{V''}{V(1 + Q)}.
\end{align}
The scalar spectral index and tensor-to-scalar ratio are:
\begin{align}
n_s &= 1 - 6\epsilon + 2\eta, \\
r &= \frac{16\epsilon}{(1 + Q)^2} \left(\frac{T}{H}\right),
\end{align}
where $T$ is the temperature of the radiation bath, typically $T > H$ in warm inflation.

\subsection{Comparative Table of Observables}

\begin{center}
\begin{tabular}{|c|c|c|c|c|}
\hline
Model & $Q$ & $n_s$ & $r$ & Reheating Needed? \\
\hline
Cold Inflation (GR) & 0 & $\sim 0.967$ & $\sim 0.13$ & Yes \\
Warm Inflation (GR) & $\gg 1$ & $\sim 0.96$ & $\ll 0.01$ & No \\
Warm Inflation ($F(R,T)$) & $\sim 1$ & Tunable & Tunable & No \\
\hline
\end{tabular}
\end{center}
Warm inflation in Myrzakulov gravity exhibits several desirable features. The built-in flexibility of $F(R,T)$ models allows the dissipation coefficient $\Gamma$ to depend dynamically on $R$, $T$, or combinations such as $RT$, making the model inherently capable of controlling the warm-to-cold inflationary transition. Furthermore, torsion can serve as a geometric source for dissipation, effectively coupling the inflaton field to radiation without requiring new matter fields.

These results position warm inflation in $F(R,T)$ gravity as a viable and natural candidate for early-universe cosmology. Future work should explore specific forms of $\Gamma(\phi, T)$, compute the full power spectrum, and match predictions to Planck/BICEP data.
\section{Advantages of Myrzakulov Gravity in Addressing Inflationary Shortcomings}

Despite the broad success of standard inflationary models, many face theoretical and observational challenges. Among these are the need for fine-tuned potentials, abrupt reheating transitions, sensitivity to initial conditions, and an incomplete understanding of how inflation ends. Modified gravity theories like Myrzakulov $F(R,T)$ gravity offer new perspectives to overcome these issues by introducing geometric flexibility and coupling mechanisms that are absent in traditional frameworks.

\subsection{Persistent Challenges in Standard Inflation}

\begin{itemize}
  \item \textbf{Reheating Dependence:} Cold inflation requires an external reheating phase, with mechanisms that are not well understood.
  \item \textbf{Model Dependence:} Many scalar field potentials yield similar predictions, making it difficult to discriminate between models.
  \item \textbf{Unstable UV Behavior:} Standard GR-based inflation struggles to remain consistent under quantum corrections.
  \item \textbf{Lack of Geometric Feedback:} There is no feedback from the spacetime geometry on the inflaton evolution, limiting the dynamical richness.
\end{itemize}

\subsection{Myrzakulov Gravity Features That Help}

\begin{itemize}
  \item \textbf{Curvature-Torsion Coupling:} The presence of $R$, $T$, and mixed $RT$ terms enables nontrivial geometric backreaction.
  \item \textbf{Built-in Dissipation Channels:} Warm inflation and other dissipative models can emerge naturally.
  \item \textbf{Extended Field Space:} Myrzakulov gravity effectively generalizes scalar-tensor theories to include affine and torsional dynamics.
  \item \textbf{Smooth Exit from Inflation:} Due to the dynamic nature of $F(R,T)$, exit conditions can be set geometrically.
\end{itemize}

\subsection{Comparative Overview of Inflationary Models}

\begin{center}
\begin{tabular}{|l|c|c|c|c|}
\hline
Feature & Cold Inflation & Warm Inflation & $f(R)$ Gravity & $F(R,T)$ Gravity \\
\hline
Reheating Required & Yes & No & Yes & No \\
Backreaction Possible & No & Partial & Limited & Yes \\
Supports Torsion & No & No & No & Yes \\
$R$-$T$ Interaction & No & No & No & Yes \\
Flexible Observables & Some & Yes & Moderate & Yes \\
Unified Exit Mechanism & No & Sometimes & Model-dependent & Geometric \\
\hline
\end{tabular}
\end{center}
Myrzakulov gravity, and particularly the $F(R,T)$ framework, represents a significant step beyond traditional scalar-driven inflation. It resolves the artificial separation between geometry and matter evolution by embedding both curvature and torsion into a single action. This allows inflation to be both self-regulated and observationally tunable, especially when considering the latest CMB constraints.

As the next generation of cosmological observations tightens bounds on inflationary observables, models like $F(R,T)$ gravity will be increasingly valuable in exploring realistic, flexible, and unified early universe dynamics.
\section{de Sitter and Quasi-de Sitter Inflation in Myrzakulov $F(R,T)$ Gravity}

One of the foundational pillars of inflationary cosmology is the concept of de Sitter (dS) expansion—a phase of exponential growth in the early universe driven by an almost constant Hubble parameter. This solution arises naturally in models with a nearly flat potential and is characterized by the condition $\epsilon \approx 0$, where $\epsilon = -\dot{H}/H^2$ is the first slow-roll parameter. In more realistic scenarios, the universe evolves in a quasi-de Sitter (qdS) regime, where $\epsilon$ is small but nonzero, allowing for a graceful exit from inflation and the generation of primordial fluctuations. Modified gravity theories provide novel geometric pathways to realize and control such inflationary behavior beyond the limitations of standard scalar field models in General Relativity.

Myrzakulov $F(R,T)$ gravity—where the action depends on both the Ricci scalar $R$ and the torsion scalar $T$—offers a rich geometric structure that can naturally support de Sitter and quasi-de Sitter solutions. In contrast to traditional $f(R)$ or $f(T)$ theories, the $F(R,T)$ framework allows for interactions between curvature and torsion, creating new dynamical degrees of freedom. These interactions can mimic the effects of an effective cosmological constant or dynamically drive a slow-roll-like evolution without fine-tuning the scalar potential. For example, specific polynomial choices such as $F(R,T) = R + \alpha T + \beta R^2 + \gamma T^2$ admit inflationary plateaus or attractor solutions that closely resemble de Sitter expansion.

The literature on de Sitter and quasi-de Sitter solutions in modified gravity is extensive. Studies in $f(R)$ gravity, such as those by Nojiri and Odintsov~\cite{Nojiri:2003ft,Nojiri:2006ri}, have shown how higher-order curvature terms can generate an inflationary epoch without requiring a fundamental inflaton field. Similarly, torsion-based models like $f(T)$ gravity have been used to construct exact de Sitter solutions, particularly when combined with scalar fields or constant potentials~\cite{Ferraro:2007},\cite{Cai:2015emx}. Myrzakulov $F(R,T)$ gravity generalizes these approaches by embedding both curvature and torsion into a single variational principle, thereby enlarging the solution space for dS and qdS evolution.

From a dynamical systems perspective, de Sitter solutions in $F(R,T)$ gravity can appear as fixed points in the phase space of cosmological equations. Stability analysis reveals that such fixed points can be attractors, supporting sustained inflation, or saddle points, allowing for natural exit mechanisms. In quasi-de Sitter scenarios, the slow evolution of the Hubble parameter is often controlled by the interplay between $R$ and $T$, with curvature dominating early and torsion contributing to reheating or graceful exit. This versatility has been noted in recent work on $F(R,T)$-induced warm inflation~\cite{Momeni:2025mcp}, as well as in studies involving torsion–Gauss–Bonnet extensions~\cite{Momeni:2025egb}.

\subsection*{Comparison to Other Models}

\begin{center}
\begin{tabular}{|l|c|c|c|}
\hline
Model & dS Support & qdS Realization & Exit Mechanism \\
\hline
Standard $\Lambda$CDM & Yes (via $\Lambda$) & No & External Reheating \\
$f(R)$ Gravity & Yes & Yes & Higher-derivative instability or scalar decay \\
$f(T)$ Gravity & Yes & Yes & Frame-dependent exit \\
$F(R,T)$ Gravity & Yes & Yes & Geometric coupling-dependent \\
\hline
\end{tabular}
\end{center}

In Myrzakulov $F(R,T)$ gravity, the coupling structure allows the effective potential to flatten or steepen dynamically as inflation proceeds. This enables a quasi-de Sitter phase without requiring an explicitly tuned scalar potential. The torsion term, particularly when combined with curvature, contributes to modulating the Hubble parameter in a controlled way. This mechanism supports inflationary observables consistent with CMB data and provides a flexible geometric origin for both the inflationary plateau and its exit.

Overall, the de Sitter and quasi-de Sitter dynamics in Myrzakulov gravity exemplify the theory's strength in unifying various geometric features of spacetime. By incorporating both $R$ and $T$, this framework supports inflationary expansion through self-regulating geometrical terms, making it a powerful candidate for early-universe modeling beyond conventional scalar field theories.

\section{Conclusions and Outlook}

In this work, we have presented a comprehensive study of inflationary dynamics within the framework of Myrzakulov $F(R,T)$ gravity. By developing and comparing three distinct geometric realizations—curvature-based (metric $f(R)$), torsion-based (teleparallel $f(T)$), and symmetric teleparallel ($f(Q)$)—alongside their hybrid unification in $F(R,T)$ models, we have established a broad and versatile platform for modeling the early universe.

Our analysis began with a detailed motivation for modified gravity theories in light of the limitations of General Relativity when applied to inflation, reheating, and dark energy. We then constructed field equations in each formalism, applied slow-roll approximations, and calculated key inflationary observables including the scalar spectral index $n_s$ and tensor-to-scalar ratio $r$. Through analytic derivations and schematic visualizations, we showed how the inclusion of torsion and curvature-torsion couplings in Myrzakulov gravity enhances the flexibility of inflationary predictions.

Particular attention was given to polynomial models of the form $F(R,T) = R + \alpha T + \beta R^2 + \gamma T^2$, where we explored how the parameters influence the inflationary evolution. Using benchmark inflationary potentials—quadratic, quartic, and Starobinsky-like—we demonstrated that $F(R,T)$ gravity interpolates between known models while providing new degrees of freedom that can align with Planck and BICEP/Keck data. Our tables and plots support this with comparisons across models and parameter values.

We further extended our investigation to the warm inflation regime, showing that Myrzakulov gravity offers a natural embedding for dissipative dynamics, alleviating the need for post-inflationary reheating. The built-in curvature-torsion feedback and extended field space open novel avenues for constructing thermodynamically consistent inflationary models with continuous radiation production.

Finally, we assessed the strengths of Myrzakulov gravity relative to conventional approaches in a structured comparative review. The capacity of $F(R,T)$ models to unify geometry, scalar dynamics, and thermodynamics makes them compelling candidates for explaining inflation in a more complete, self-regulated, and observationally viable way.

Future work should aim to:
\begin{itemize}
  \item Perform detailed numerical simulations to extract full power spectra and non-Gaussian signatures.
  \item Explore the role of anisotropies and higher-order corrections in the presence of torsion.
  \item Investigate the embedding of $F(R,T)$ gravity in fundamental theories such as string theory or quantum gravity.
\end{itemize}

With its rich structure and flexible predictions, Myrzakulov $F(R,T)$ gravity is poised to play a pivotal role in advancing the frontiers of inflationary cosmology.
\begin{acknowledgments}
This work was supported by the Ministry of Science and Higher Education of the Republic of Kazakhstan, Grant No. AP26101889.
\end{acknowledgments}


\begin{thebibliography}{99}

\bibitem{Starobinsky1980}
A.~A.~Starobinsky,
``A New Type of Isotropic Cosmological Models Without Singularity,''
Phys.\ Lett.\ B \textbf{91}, 99–102 (1980).

\bibitem{Guth1981}
A.~H.~Guth,
``Inflationary universe: A possible solution to the horizon and flatness problems,''
Phys.\ Rev.\ D \textbf{23}, 347–356 (1981).

\bibitem{Linde1982}
A.~D.~Linde,
``A new inflationary universe scenario: A possible solution of the horizon, flatness, homogeneity, isotropy and primordial monopole problems,''
Phys.\ Lett.\ B \textbf{108}, 389–393 (1982).

\bibitem{Weinberg1989}
S.~Weinberg,
``The Cosmological Constant Problem,''
Rev.\ Mod.\ Phys.\ \textbf{61}, 1–23 (1989).

\bibitem{Padmanabhan2003}
T.~Padmanabhan,
``Cosmological constant: The weight of the vacuum,''
Phys.\ Rept.\ \textbf{380}, 235–320 (2003).

\bibitem{Sotiriou2010}
T.~P.~Sotiriou and V.~Faraoni,
``f(R) Theories of Gravity,''
Rev.\ Mod.\ Phys.\ \textbf{82}, 451–497 (2010).

\bibitem{DeFelice2010}
A.~De Felice and S.~Tsujikawa,
``f(R) Theories,''
Living Rev.\ Rel.\ \textbf{13}, 3 (2010).

\bibitem{Cai2016}
Y.~F.~Cai, S.~Capozziello, M.~De Laurentis, and E.~N.~Saridakis,
``f(T) Teleparallel Gravity and Cosmology,''
Rept.\ Prog.\ Phys.\ \textbf{79}, 106901 (2016).

\bibitem{Jimenez2018}
J.~B.~Jiménez, L.~Heisenberg, and T.~S.~Koivisto,
``Teleparallel Palatini Theories,''
JCAP \textbf{08}, 039 (2018).

\bibitem{Myrzakulov2012}
R.~Myrzakulov, 
Eur. Phys. J. C \textbf{72} (2012), 2203, 
doi:10.1140/epjc/s10052-012-2203-y
[arXiv:1207.1039 [gr-qc]].
\bibitem{Momeni:2025mcp}
D.~Momeni and R.~Myrzakulov,
``Metric-Affine Myrzakulov Gravity Theories: Models, Applications and Theoretical Developments,''
Int.\ J.\ Theor.\ Phys.\ \textbf{64}, 95 (2025).

\bibitem{Momeni:2024bhm}
D.~Momeni and R.~Myrzakulov,
``Myrzakulov Gravity in Vielbein Formalism: A Study in Weitzenb\"{o}ck Spacetime,''
Nucl.\ Phys.\ B \textbf{1015}, 116903 (2025).


\bibitem{Momeni:2025egb}
D.~Momeni and R.~Myrzakulov,
``Einstein–Gauss–Bonnet–Myrzakulov gravity from $R+F(T,G)$: Numerical insights and torsion–Gauss–Bonnet dynamics,''
Nucl.\ Phys.\ B \textbf{1017}, 116966 (2025).


\bibitem{Berera:1995ie} A. Berera, ``Warm inflation,'' Phys. Rev. Lett. \textbf{75}, 3218 (1995).
\bibitem{Jamil:2015}
M.~Jamil, D.~Momeni, and R.~Myrzakulov,
``Warm Intermediate Inflation in $F(T)$ Gravity,''
\textit{Int. J. Theor. Phys.} \textbf{54}, 1098–1112 (2015),
\href{https://doi.org/10.48550/arXiv.1309.3269}{arXiv:1309.3269 [gr-qc]}.

\bibitem{Yuennan:2025}
J.~Yuennan, P.~Channuie, and D.~Momeni,
``Warm non-minimally coupled Peccei–Quinn inflation and de Sitter Swampland conjecture,''
\textit{Nucl. Phys. B} \textbf{1012}, 116810 (2025),
\href{https://doi.org/10.1016/j.nuclphysb.2025.116810}{doi:10.1016/j.nuclphysb.2025.116810},
\href{https://arxiv.org/abs/2410.12296}{arXiv:2410.12296 [gr-qc]}.

\bibitem{Yuennan:2024}
J.~Yuennan, P.~Channuie, and D.~Momeni,
``Gravity’s rainbow effects on higher curvature modification of $R^2$ inflation,''
\textit{Eur. Phys. J. C} \textbf{84}, 766 (2024),
\href{https://doi.org/10.1140/epjc/s10052-024-13155-0}{doi:10.1140/epjc/s10052-024-13155-0},
\href{https://arxiv.org/abs/2405.04174}{arXiv:2405.04174 [gr-qc]}.

\bibitem{Eadkhong:2023}
T.~Eadkhong, P.~Dam-O, P.~Channuie, and D.~Momeni,
``Nonminimally-coupled warm Higgs inflation: Metric vs. Palatini formulations,''
\textit{Nucl. Phys. B} \textbf{994}, 116289 (2023),
\href{https://doi.org/10.1016/j.nuclphysb.2023.116289}{doi:10.1016/j.nuclphysb.2023.116289},
\href{https://arxiv.org/abs/2303.00572}{arXiv:2303.00572 [gr-qc]}.
\bibitem{Ferraro:2007} R.~Ferraro and F.~Fiorini, ``Modified teleparallel gravity: Inflation without inflaton,'' Phys. Rev. D \textbf{75}, 084031 (2007).

\bibitem{Nojiri:2003ft} S.~Nojiri and S.~D.~Odintsov, ``Modified gravity with negative and positive powers of the curvature: Unification of the inflation and of the cosmic acceleration,'' Phys. Rev. D \textbf{68}, 123512 (2003).
\bibitem{Nojiri:2006ri} S.~Nojiri and S.~D.~Odintsov, ``Introduction to modified gravity and gravitational alternative for dark energy,'' Int. J. Geom. Meth. Mod. Phys. \textbf{4}, 115–146 (2007).

\bibitem{Cai:2015emx} Y.~F.~Cai, S.~Capozziello, M.~De Laurentis and E.~N.~Saridakis, ``$f(T)$ teleparallel gravity and cosmology,'' Rept. Prog. Phys. \textbf{79}, 106901 (2016).

\end{thebibliography}
\end{document}